\documentclass[english,aps,preprint]{revtex4}
\usepackage{times}
\usepackage[T1]{fontenc}
\usepackage[latin1]{inputenc}
\usepackage{amsmath}
\usepackage{color}
\usepackage{graphicx}
\usepackage{amssymb}

\makeatletter
\usepackage{babel}
\makeatother
\begin{document}

\preprint{This line only printed with preprint option}

\title{Dispersive destabilization of nonlinear light propagation in fiber
Bragg gratings: a numerical verification}

\author{Carlos Martel}

\email{martel@fmetsia.upm.es}

\author{Carlos M. Casas}

\affiliation{Depto. de Fundamentos Matemáticos, E.T.S.I. Aeronáuticos, Universidad
Politécnica de Madrid, Plaza Cardenal Cisneros 3, 28040 Madrid, Spain }

\begin{abstract}
This paper presents some numerical simulations of the full one-dimensional
Maxwell-Lorentz equations that describe light propagation in fiber
Bragg gratings in order to confirm that the standard nonlinear coupled
mode equations fail to predict the weakly nonlinear dynamics of the
system when dispersive instabilities come into play, and that, in
this case, the correct slow envelope description of the system requires
to consider higher order dispersion effects.
\end{abstract}
\maketitle

\section{introduction}

\textbf{The nonlinear coupled mode equations (NLCME) are the envelope
equations currently used to study the weakly nonlinear dynamics of
light propagation in fiber Bragg gratings. These equations do not
include dispersion effects. In this paper we integrate numerically
the full 1D Maxwell-Lorentz equations in a fiber grating in order
to show that the dispersion effects can be essential in the dynamics
of the system and that the correct weakly nonlinear description of
the system has necessarily to include higher order dispersion terms.
The resulting envelope equations are asymptotically nonuniform in
the sense that they include terms with different asymptotic order,
and this is a standard situation for general extended, propagative
(i.e., with order one group velocity) pattern forming systems.}

The weakly nonlinear dynamics of resonant light propagation in a Fiber
Bragg grating (FBG), i.e., optical fiber with a periodic variation
of the refractive index along its length, is usually described using
the so-called nonlinear coupled mode equations (NLCME)\textcolor{black}{\begin{eqnarray}
 & A_{t}^{+}=\phantom{-}A_{x}^{+}+\textrm{i}\kappa A^{-}+\textrm{i}A^{+}(\sigma|A^{+}|^{2}+|A^{-}|^{2}),\label{NLCME1}\\
 & A_{t}^{-}=-A_{x}^{-}+\textrm{i}\kappa A^{+}+\textrm{i}A^{-}(\sigma|A^{-}|^{2}+|A^{+}|^{2}),\label{NLCME2}\end{eqnarray}
which prescribe the evolution of the complex envelopes $A^{\pm}$
of the two slowly modulated resonant wavetrains that approximately
constitute the actual field inside the FBG} \begin{equation}
E\sim\begin{array}{c}
A^{+}(x,t)\end{array}\textrm{e}^{\textrm{i}x+\textrm{i}\omega t}+A^{-}(x,t)\textrm{e}^{-\textrm{i}x+\textrm{i}\omega t}+\:\mbox{c.c.}+\cdots,\label{Efield}\end{equation}
 see e.g. \textcolor{black}{\cite{WinfulCooperman82,deSterkeSipe94,deSterke98,Aceves00,GoodmanWeinsteinHolmes01}}.
The NLCME above, where space, time and the amplitudes have been rescaled
to reduce the number of parameters, retain the combined effect of
the group velocity, the coupling induced by the grating and the weakly
nonlinear interaction of the wavetrains. This formulation, apart from
FBG, has been also used to describe the evolution of quasi-onedimensional
Bose-Einstein condensates in optical lattices \cite{yulinskryabin03,SakaguchiMalomed04}
and, in general, the NLCME are commonly regarded as the normal form
for the weakly nonlinear dynamics of any extended, propagative system
without dissipation and with a weak spatial periodic structure.

In a recent paper \cite{Martel05} one of the authors showed that
the NLCME (\ref{NLCME1})-(\ref{NLCME2}) fail to predict the dynamics
of the system when dispersive instabilities (that cannot be detected
using the NLCME formulation) come into play, and that, for both signs
of the dispersion coefficient, there are always stable solutions according
to the NLCME that are dispersively unstable. In order to correctly
describe the weakly nonlinear evolution of the system, the effect
of higher order dispersion has to be retained and the appropriate
amplitude equations are the following dispersive nonlinear coupled
mode equations (NLCMEd\textcolor{black}{)} \textcolor{black}{\begin{eqnarray}
 & A_{t}^{+}=\phantom{-}A_{x}^{+}+\textrm{i}\kappa A^{-}+\textrm{i}A^{+}(\sigma|A^{+}|^{2}+|A^{-}|^{2})+\textrm{i}\varepsilon A_{xx}^{+},\label{disp+}\\
 & A_{t}^{-}=-A_{x}^{-}+\textrm{i}\kappa A^{+}+\textrm{i}A^{-}(\sigma|A^{-}|^{2}+|A^{+}|^{2})+\textrm{i}\varepsilon A_{xx}^{-}.\label{disp-}\end{eqnarray}
}

\textcolor{black}{This dispersive system was introduced and analyzed
in detail in} \cite{Martel05}, but we think it is convenient to briefly
remind here some of the results obtained in that paper: 

\begin{enumerate}
\item The scaling of the NLCMEd is the same of the NLCME: the characteristic
length scale is the slow scale that results from the balance of the
advection term with the small effect of the grating, the characteristic
time is the corresponding transport time scale and the characteristic
size of the wavetrains results from the saturation of the small nonlinear
terms. The small amplitude slow envelope assumption, which is the
key assumption that allow us to derive both systems of equations,
forces the dispersive terms to be always small as compared with the
advection terms (in other words, the nonzero group velocity turns
this system into a transport dominated one) and therefore (with the
scaling mentioned above) the NLCMEd must be considered only in the
physically relevant regime $\varepsilon\rightarrow0$.
\item The NLCMEd are asymptotically nonuniform, in the sense that the resulting
asymptotic model, in the $\varepsilon\rightarrow0$ limit, still contains
the small parameter $\varepsilon$. This is due to the fact that the
NLCMEd include simultaneously two balances with different asymptotic
order: one induced by the dominant transport terms and the other associated
with the underlying effect of dispersion. This kind of asymptotically
nonuniform amplitude equations have been previously derived in the
context of water waves \cite{martelvegaknobloch03} and for the onset
of the oscillatory instability in spatially extended dissipative systems
\cite{martelvega96,martelvega98}. 
\item Two spatial scales are present in the NLCMEd: transport scales $\delta x_{\textrm{trans}}\sim1$,
and dispersive scales $\delta x_{\textrm{disp}}\sim\sqrt{|\varepsilon|}\ll1$.
The dispersive scales are small as compared with the transport scales
but still large as compared with the wavelength of the basic resonant
wavetrains in expression (\ref{Efield}), which, in the scaling we
are using, is of the order of $|\varepsilon|\ll1$, and therefore
the slow envelope assumption is not violated.
\item If only transport scales are present, then the dispersion terms in
the NLCMEd $|\varepsilon A_{xx}|,|\varepsilon B_{xx}|\sim|\varepsilon|\ll1$
can be safely neglected (they produce only a small quantitative correction
that vanishes as $\epsilon\rightarrow0$) and the evolution of the
system is well represented by the NLCME. On the other hand, if the
small dispersive scales do develop, then the NLCME do not correctly
predict the dynamics system and the NLCMEd must be used instead. The
onset of the dispersive scales is not a higher order, longer time
effect; it takes place in the same timescale of the NLCME no matter
how small the dispersion coefficient $\epsilon$ is. Once the dispersive
scales appear they typically spread all over the domain giving rise
to very complicated spatio-temporal dynamics. This dispersive destabilization
can be simply regarded as the standard modulational instability of
the NLS-like dynamics that lays beneath the dominant transport induced
dynamics.
\item The stability of the family of uniform modulus solutions, known as
continuous waves (CW), is drastically affected by dispersion. The
stability predictions for the CW from the NLCME differ completely
from those obtained from the NLCMEd for both signs of the dispersion
coefficient $\epsilon$, no matter how small it may be.
\end{enumerate}
Despite of the results presented in \cite{Martel05} it appears that
the NLCME continue to be used as the amplitude equations for the description
of light propagation in FBG and for the weakly nonlinear dynamics
of BEC in optical lattices without paying any attention to the effect
of dispersion. In order to make clear that the correct amplitude equations
are the NLCMEd, we have decided to carry out some numerical integrations
of the full 1D Maxwell-Lorenz equations (MLE) in a long fiber Bragg
grating and check that the stability predictions for the CW given
by the NLCME are wrong and that the dispersive NLCMEd give the correct
results.

This paper is organized as follows: in the following section we derive
the explicit expressions of the coefficients of the NLCMEd from the
MLE and, in the next and final section of this paper, we present some
numerical integrations of the MLE starting from a perturbed CW and
compare them with the CW stability characteristics predicted by the
NLCME and the NLCMEd.

\section{NLCMEd derivation from the MLE}

Our formulation follows closely that of ref. \cite{GoodmanWeinsteinHolmes01}.
We have decided to include here a quited detailed derivation of the
NLCMEd from the MLE because we use rather new derivation procedure
that has the advantage of not requiring to assume any a priori relation
among the different small parameters of the problem. 

\begin{figure*}[h]
\begin{centering}\includegraphics[bb=0bp 0bp 435bp 209bp,clip,scale=0.5]{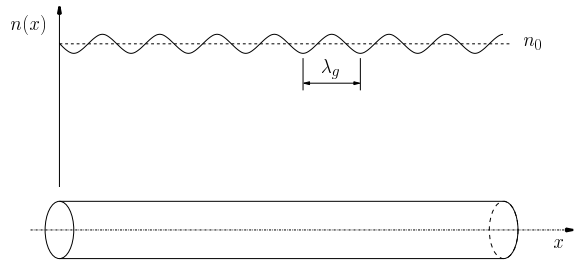}\par\end{centering}

\caption{\label{fig:FBG}One dimensional fiber with a periodic variation of
the refractive index.}
\end{figure*}

We describe the propagation of light in a fiber with a periodic grating
and a cubic nonlinearity using the one-dimensional Maxwell's equations
\cite{Agrawal95,LauterbornKW93} for the evolution of the electromagnetic
fields together with an anharmonic Lorentz oscillator model for the
polarization (see e.g. \cite{GoodmanWeinsteinHolmes01,SorensenBWM02}
and references therein) \begin{eqnarray}
 & \dfrac{\partial B}{\partial t}=\dfrac{\partial E}{\partial x},\label{MLdim1}\\
 & \mu_{0}\dfrac{\partial D}{\partial t}=\dfrac{\partial B}{\partial x},\label{MLdim2}\\
 & D=\epsilon_{0}E+P,\label{MLdim3}\\
 & \Omega_{p}^{-2}\dfrac{\partial^{2}P}{\partial t^{2}}+(1-2\Delta n\cos(2\pi x/\lambda_{g}))P-\gamma P^{3}=\epsilon_{0}\chi E.\label{MLdim4}\end{eqnarray}
In the system above, the electric field $E,$ the magnetic field $B$,
the dielectric displacement $D$ and the polarization $P$ are scalar
fields that depend on the spatial variable $x$ and on time $t.$
$\mu_{0}$ and $\epsilon_{0}$ denote, respectively, the permeability
and the permittivity of the vacuum. The characteristic frequency $\Omega_{p}$
accounts for the non instantaneous polarization response of the media,
$\Delta n$ and $\lambda_{g}$ represent the strength and the period
of the grating, that is, the strength and the period of the spatial
periodic variation of the refractive index of the fiber ($\Delta n$
measures the size of the nonuniformities of the refraction index relative
to its mean value $n_{0}$, see Fig.~\ref{fig:FBG}), $\chi$ is
the linear polarizability of the medium ($n_{0}^{2}=1+\chi$) and
$\gamma>0$ is the coefficient of the nonlinear Kerr effect.

In order to simplify subsequent calculations it is convenient to make
the system (\ref{MLdim1})-(\ref{MLdim4}) nondimensional using the
following rescalings:\begin{eqnarray*}
 & B=\sqrt{\mu_{0}/(\epsilon_{0}\gamma)}\tilde{B,}\quad D=(1/\sqrt{\gamma}\tilde{)D,}\quad E=(1/\sqrt{(\epsilon_{0}\gamma)}\tilde{)E,}\quad P=(1/\sqrt{\gamma})\tilde{P},\\
 & x=(\lambda_{g}/\pi)\tilde{x},\quad t=(\lambda_{g}/c\pi)\tilde{t,}\end{eqnarray*}
where $c^{2}=1/(\epsilon_{0}\mu_{0})$ is the vacuum speed of light.
After dropping tildes and eliminating $D$ and $B$, the nondimensional
MLE can be written in the form \begin{eqnarray}
 &  & \dfrac{\partial^{2}(E+P)}{\partial t^{2}}=\dfrac{\partial^{2}E}{\partial x^{2}},\label{EP1}\\
 &  & \dfrac{\partial^{2}P}{\partial t^{2}}=-\omega_{p}^{2}(1-2\Delta n\cos(2x))P+\omega_{p}^{2}(n_{0}^{2}-1)E+\omega_{p}^{2}P^{3}.\label{EP2}\end{eqnarray}
where the grating period is now equal to $\pi$ and the dimensionless
finite time polarization response frequency is given by $\omega_{p}^{2}=\Omega_{p}^{2}\lambda_{g}^{2}/(c^{2}\pi^{2})$.

\begin{figure*}[h]
\begin{centering}\includegraphics[clip,scale=0.5]{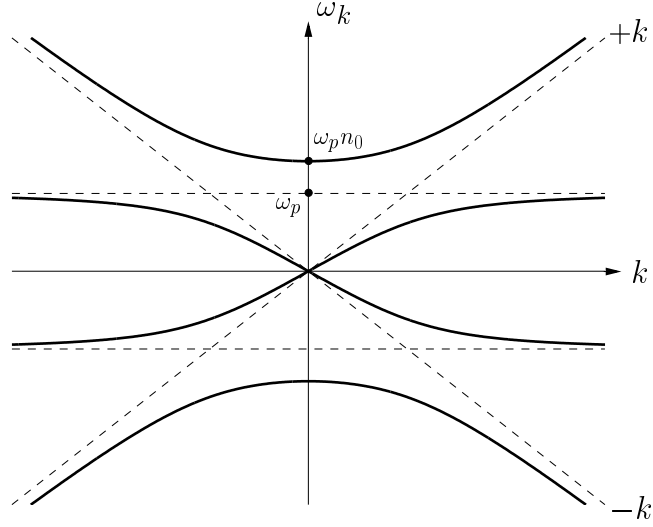}\par\end{centering}

\caption{\label{fig:disp-rel}Sketch of the dispersion relation (\ref{disprel}).}
\end{figure*}

In the absence of grating, the linear propagation characteristics
of a wavetrain of the form\begin{equation}
\begin{array}{c}
\left\{ \begin{array}{c}
E(x,t)\\
P(x,t)\end{array}\right\} =\left\{ \begin{array}{c}
E_{k}\\
P_{k}\end{array}\right\} \end{array}e^{ikx+i\omega_{k}t}+\:\mbox{c.c.},\label{WTlineal}\end{equation}
are given by the following dispersion relation\begin{equation}
\omega_{k}^{4}-\omega_{k}^{2}(k²+\omega_{p}^{2}n_{0}^{2})+\omega_{p}^{2}k^{2}=0,\label{poli}\end{equation}
which, for $n_{0}^{2}>1$, has four real roots of the form \begin{equation}
\omega_{k}=\pm\sqrt{(k²+\omega_{p}^{2}n_{0}^{2})/2\pm\sqrt{(k²+\omega_{p}^{2}n_{0}^{2})^{2}/4-\omega_{p}^{2}k^{2}}},\label{disprel}\end{equation}
and associated eigenvectors\begin{equation}
\begin{array}{c}
\left\{ \begin{array}{c}
E_{k}\\
P_{k}\end{array}\right\} =\left\{ \begin{array}{c}
\omega_{k}^{2}\\
k^{2}-\omega_{k}^{2}\end{array}\right\} \end{array}.\label{autovect}\end{equation}
The four branches of the dispersion relation (\ref{disprel}) are
plotted in Fig.~\ref{fig:disp-rel}. There are two different behaviors
for large wavenumbers: one is dominated by the finite time polarization
response of the medium, $\omega_{k}\rightarrow\pm\omega_{p}$ as $k\rightarrow\pm\infty$,
and the other, $\omega_{k}\rightarrow\pm k$ as $k\rightarrow\pm\infty$,
corresponds to propagation like in the vacuum, without polarization
effects.

The small nonuniformities of the refractive index, $\Delta n\ll1$,
and the effect of the small nonlinearity can be accounted for by allowing
the wavetrains that resonate with the grating to be slowly modulated
in space and time \begin{equation}
\left\{ \begin{array}{c}
E(x,t)\\
P(x,t)\end{array}\right\} =V_{0}(A^{+}(x,t)e^{ix+i\omega t}+A^{-}(x,t)e^{-ix+i\omega t})+\:\mbox{c.c.}\:+\dots,\label{ansatz}\end{equation}
where\begin{equation}
V_{0}=\begin{array}{c}
\left\{ \begin{array}{c}
\omega^{2}\\
1-\omega^{2}\end{array}\right\} \end{array},\qquad\mbox{and}\qquad\omega=\sqrt{(1+\omega_{p}^{2}n_{0}^{2})/2\pm\sqrt{(1+\omega_{p}^{2}n_{0}^{2})^{2}/4-\omega_{p}^{2}}},\label{eigv+omega}\end{equation}
and the weakly nonlinear level of this approach requires essentially
that\begin{equation}
\dots\ll|A_{xx}^{\pm}|\ll|A_{x}^{\pm}|\ll|A^{\pm}|\ll1,\quad\dots\ll|A_{t}^{\pm}|\ll|A^{\pm}|\ll1\quad\mbox{and}\quad\Delta n\ll1,\label{smallslow}\end{equation}
 that is, small amplitudes that depend slowly on space and time and
small grating strength. The solution of eqs. (\ref{EP1})-(\ref{EP2})
and the amplitude equations can be expanded in powers of the small
quantities $\Delta n$, $A^{\pm}$, $A_{x}^{\pm}$, $A_{xx}^{\pm}$,$\ldots$
as \begin{eqnarray}
 & \left\{ \begin{array}{c}
E(x,t)\\
P(x,t)\end{array}\right\} =V_{0}(A^{+}e^{ix+i\omega t}+A^{-}e^{-ix+i\omega t})+\:\mbox{c.c.}+\nonumber \\
 & \qquad\qquad\qquad\qquad\qquad+v_{1}^{+}A_{x}^{+}+v_{1}^{-}A_{x}^{-}+v_{2}^{+}A_{xx}^{+}+v_{2}^{-}A_{xx}^{-}+\dots,\label{expanSOL}\\
 & A_{t}^{+}=\alpha_{0}^{+}A^{+}+\alpha_{1}^{+}A_{x}^{+}+\alpha_{2}^{+}A_{xx}^{+}+\dots,\label{AE+}\\
 & A_{t}^{-}=\alpha_{0}^{-}A^{-}+\alpha_{1}^{-}A_{x}^{-}+\alpha_{2}^{-}A_{xx}^{-}+\dots,\label{AE-}\end{eqnarray}
which, once inserted into eqs. (\ref{EP1})-(\ref{EP2}), provide
a linear nonhomogeneous system for the contribution of each order.
For the resonant terms, i.e., those proportional to $e^{\pm ix\pm i\omega t}$,
a condition must be satisfied to ensure that there are not secular
terms in the short time scale. In other words, the linear problems
corresponding to the resonant terms are singular and hence a solvability
condition must be satisfied by the nonhomogeneous part; these solvability
conditions yield the coefficients of the amplitude equations.

Notice that only the resonant terms contribute to the amplitude equations
and only the amplitude equation for $A^{+}$ has to be calculated
because the corresponding equation for $A^{-}$ can be obtained by
simply applying the symmetry\begin{equation}
x\rightarrow-x\qquad A^{+}\longleftrightarrow A^{-},\label{spatialsymm}\end{equation}
which comes from the spatial reflection symmetry of the original problem
(\ref{EP1})-(\ref{EP2}). 

The linear terms in the amplitude equations can be easily anticipated
because they correspond to the Taylor expansion of the dispersion
relation (\ref{disprel}) at $k=1$ (see e.g. \cite{crosshohenberg93}),\[
\left.i(\omega_{k}\right|_{k=1}-\omega)A^{+}+\left.\frac{d\omega_{k}}{dk}\right|_{k=1}A_{x}^{+}-i\frac{1}{2}\left.\frac{d^{2}\omega_{k}}{dk^{2}}\right|_{k=1}A_{xx}^{+}+\dots.\]
The first coefficient obviously vanishes (see eq. (\ref{eigv+omega}))
and the second and third coefficients are, respectively, the group
velocity and the higher order dispersion, which, after making use
of eq. (\ref{poli}), can be written as\begin{eqnarray}
 &  & v_{g}=\left.\frac{d\omega_{k}}{dk}\right|_{k=1}=\frac{\omega(\omega^{2}-\omega_{p}^{2})}{\omega^{4}-\omega_{p}^{2}},\label{velgrupo}\\
 &  & id=-i\frac{1}{2}\left.\frac{d^{2}\omega_{k}}{dk^{2}}\right|_{k=1}=-i\frac{1}{2}\frac{\omega^{3}(\omega^{2}-1)(\omega^{2}-\omega_{p}^{2})(3\omega_{p}^{2}+\omega^{4})}{(\omega^{4}-\omega_{p}^{2})^{3}},\label{dispersion}\end{eqnarray}
which correspond to the group velocity and dispersion of the fiber
without grating.

The first order, resonant contributions of the grating to the expansion
of the solution (\ref{ansatz}) and to the amplitude equation (\ref{AE+})
are of the form\[
W\Delta nA^{-}e^{ix+i\omega t}\qquad\mbox{and}\qquad w\Delta nA^{-},\]
where the two component vector $W$ is given by the following linear,
singular nonhomogeneous problem\[
\left[\begin{array}{cc}
\omega^{2}-1 & \omega^{2}\\
(n_{0}^{2}-1)\omega_{p}^{2} & \omega^{2}-\omega_{p}^{2}\end{array}\right]W=-\omega_{p}^{2}\left[\begin{array}{cc}
0 & 0\\
0 & 1\end{array}\right]V_{0}+2i\omega w\left[\begin{array}{cc}
1 & 1\\
0 & 1\end{array}\right]V_{0}.\]
This system can be solved only if the right hand side is orthogonal
to the solution of the adjoint problem\[
V_{0}^{a}=\begin{array}{c}
\left\{ \begin{array}{c}
\omega_{p}^{2}-\omega^{2}\\
\omega^{2}\end{array}\right\} \end{array},\]
and this solvability condition gives the value of the coefficient
of the amplitude equation\begin{equation}
w=i\frac{\omega(1-\omega^{2})}{2(\omega^{4}-\omega_{p}^{2})}\omega_{p}^{2}.\label{grating}\end{equation}

The first order contributions of the nonlinear term,\[
U_{1}A^{+}|A^{+}|^{2}e^{ix+i\omega t},\: U_{2}A^{+}|A^{-}|^{2}e^{ix+i\omega t}\qquad\mbox{and}\qquad u_{1}A^{+}|A^{+}|^{2},\: u_{2}A^{+}|A^{-}|^{2},\]
are computed similarly: the following linear problems are obtained
for the vectors $U_{1}$ and $U_{2}$\begin{eqnarray*}
\left[\begin{array}{cc}
\omega^{2}-1 & \omega^{2}\\
(n_{0}^{2}-1)\omega_{p}^{2} & \omega^{2}-\omega_{p}^{2}\end{array}\right]U_{1}=-3\omega_{p}^{2}\left[\begin{array}{c}
0\\
(1-\omega^{2})^{3}\end{array}\right]+2i\omega u_{1}\left[\begin{array}{cc}
1 & 1\\
0 & 1\end{array}\right]V_{0},\\
\left[\begin{array}{cc}
\omega^{2}-1 & \omega^{2}\\
(n_{0}^{2}-1)\omega_{p}^{2} & \omega^{2}-\omega_{p}^{2}\end{array}\right]U_{2}=-6\omega_{p}^{2}\left[\begin{array}{c}
0\\
(1-\omega^{2})^{3}\end{array}\right]+2i\omega u_{2}\left[\begin{array}{cc}
1 & 1\\
0 & 1\end{array}\right]V_{0},\end{eqnarray*}
 and, after applying the solvability condition, the resulting amplitude
equation coefficients are given by \begin{equation}
u_{1}=i\frac{3\omega(1-\omega^{2})^{3}}{2(\omega^{4}-\omega_{p}^{2})}\omega_{p}^{2}\quad\mbox{and}\quad u_{2}=i\frac{3\omega(1-\omega^{2})^{3}}{(\omega^{4}-\omega_{p}^{2})}\omega_{p}^{2}.\label{nonlinear}\end{equation}
The ratio $u_{2}=2u_{1}$could have been advanced; it is a well known
result of the cubic nonlinearity of the problem \cite{crosshohenberg93}. 

Collecting the coefficients above (\ref{velgrupo})-(\ref{nonlinear})
and applying the spatial reflection symmetry (\ref{spatialsymm})
the resulting amplitude equations can be written as\begin{eqnarray}
A_{t}^{+}=\phantom{-}v_{g}A_{x}^{+}+idA_{xx}^{+}+w\Delta nA^{-}+A^{+}(u_{1}|A^{+}|^{2}+u_{2}|A^{-}|^{2})+\dots,\label{Anoesc+}\\
A_{t}^{-}=-v_{g}A_{x}^{-}+idA_{xx}^{-}+w\Delta nA^{+}+A^{-}(u_{1}|A^{-}|^{2}+u_{2}|A^{+}|^{2})+\dots.\label{Anoesc-}\end{eqnarray}
It is important to emphasize that no particular scaling among the
small size of the amplitudes, the small grating depth, the slow time
and the large spatial scale has been used; only the slow envelope,
weakly nonlinear assumption expressed in (\ref{smallslow}) is actually
required to obtain the above amplitude equations. 

We will consider the simplest possible geometrical configuration:
propagation of light in a fiber ring with length $L\gg1$. The spatial
periodicity condition implies that the boundary conditions for $A^{+}$
and $A^{-}$ are (see eq. (\ref{ansatz}))\begin{equation}
A^{+}(x+L)e^{i\theta}=A^{+}(x,t),\quad A^{-}(x+L)e^{-i\theta}=A^{-}(x,t).\label{perodicityteta}\end{equation}
Here $\theta=L\,(\mbox{mod}2\pi)$ measures the mismatch between the
natural wavelength of the resonant wavetrains (=$2\pi)$ and the period
of the domain, but we will confine ourselves to the particular case
$\theta=0$, i.e., ring length equals to an integer multiple of the
period of the wavetrains.

\begin{figure*}[h]
\begin{centering}\includegraphics[clip,scale=0.5]{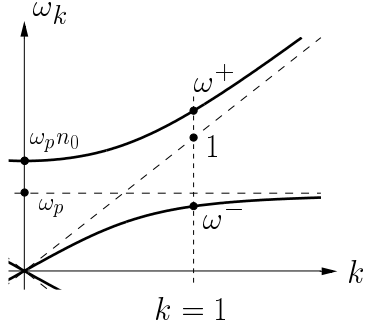}\par\end{centering}

\caption{\label{fig:disp-rel-detail}Detail of the dispersion relation (\ref{disprel})
with the two frequencies $\omega^{\pm}$ for $k=1$.}
\end{figure*}

There are two possible choices $\omega^{\pm}$ depending on the sign
selected in (\ref{eigv+omega}), see Fig.~\ref{fig:disp-rel-detail}.
The group velocity (\ref{velgrupo}) is positive in both cases (it
is the slope of the curve $\omega_{k}$ at $k=1$ in Fig.~\ref{fig:disp-rel-detail}),
but the sign of the dispersion coefficient $d$ (\ref{dispersion}),
which is related to the curvature of the curve $\omega_{k}$ in Fig.~\ref{fig:disp-rel-detail},
changes. On the other hand, the nonlinear and grating terms have imaginary
parts that are always negative; see eqs. (\ref{grating}) and (\ref{nonlinear})
and Fig.~\ref{fig:disp-rel-detail}, and recall that, using the dispersion
relation eq. (\ref{poli}) for $k=1$, the denominator can be written
as $\omega^{4}-\omega_{p}^{2}=\omega^{2}[(\omega^{2}-1)+(\omega^{2}-\omega_{p}^{2}n_{0}^{2})]$.

In order to make the nonlinear and grating coefficients positive,
we will work with the complex conjugates of the amplitudes and, to
absorb some parameters of the problem, we will also perform the following
rescalings \begin{equation}
x=L\tilde{x},\quad t=(L/v_{g})\tilde{t},\quad\overline{A^{\pm}}=\sqrt{v_{g}/(L|u_{2}|)}\tilde{A^{\pm}},\label{eq:escaladosA+-}\end{equation}
that, after dropping tildes, yield the scaled NLCMEd\begin{eqnarray}
A_{t}^{+}=\phantom{-}A_{x}^{+}+i\varepsilon A_{xx}^{+}+i\kappa A^{-}+iA^{+}(\sigma|A^{+}|^{2}+|A^{-}|^{2}),\label{A+disp}\\
A_{t}^{-}=-A_{x}^{-}+i\varepsilon A_{xx}^{-}+i\kappa A^{+}+iA^{-}(\sigma|A^{-}|^{2}+|A^{+}|^{2}),\label{A-disp}\\
A^{\pm}(x+1,t)=A^{\pm}(x,t),\qquad\qquad\qquad\label{CC-disp}\end{eqnarray}
where $\varepsilon=-d/(Lv_{g})\ll1$ is positive (negative) for $\omega=\omega^{+}$($\omega=\omega^{-}$),
the scaled grating strength $\kappa=\Delta nL|w|/v_{g}\sim1$ is always
positive, and the nonlinear coefficient $\sigma=\frac{1}{2}$ (the
standard NLCME are obtained by just by setting $\varepsilon=0$ in
the system above).

\section{Numerical results}

In order to confirm that the correct stability predictions for the
MLE are those given by the NLCMEd, we numerically integrate the complete
MLE (\ref{EP1})-(\ref{EP2}) in a large ring shaped fiber grating,
that is, with periodic boundary conditions,\[
E(x+L,t)=E(x,t),\quad P(x+L,t)=P(x,t),\]
and $L\gg1$. The MLE are integrated numerically as a system of four
first order equations, using Fourier series in space and a 4th order
Runge-Kutta scheme \cite{Lambert95} for the time integration of the
resulting ODEs. The linear diagonal terms are integrated implicitly
and the nonlinear terms are computed in physical space using the 2/3
rule to remove the aliasing terms \cite{CanutoHussaniQuarteroniZang}.
The number of modes used in the simulations presented is $M_{\text{Fourier}}=1024$
and the time step $\Delta t=.01$, and the Fourier transforms were
performed using the FFTW routines \cite{FrigoJohnson99}.

The initial condition for all simulations is a CW \textcolor{black}{\begin{eqnarray*}
 & A_{\textrm{cw}}^{+}=\rho\cos\theta\:\textrm{e}^{\textrm{i}\alpha t+\textrm{i}mx},\quad A_{\textrm{cw}}^{-}=\rho\sin\theta\:\textrm{e}^{\textrm{i}\alpha t+\textrm{i}mx},\\
 & \alpha=\dfrac{\kappa}{\sin2\theta}+\dfrac{1+\sigma}{2}\rho,\quad m=(\dfrac{\kappa}{\sin2\theta}+\dfrac{1-\sigma}{2}\rho^{2})\cos2\theta,\end{eqnarray*}
where} $\rho>0$ is the light intensity in the fiber and $\theta\in]-\frac{\pi}{2},0[\cup]0,\frac{\pi}{2}[$
measures the ratio between the two counterpropagating wavetrains (see
\cite{Martel05}), with a small superimposed perturbation. Once a
CW has been selected ($\kappa$, $\rho$ and $\theta$ fixed) and
the three MLE parameters $\omega_{p}^{2}$, $n_{0}^{2}$ and $L$
are prescribed, the initial condition for the MLE is obtained from
\begin{equation}
\left\{ \begin{array}{c}
E\\
P\end{array}\right\} =\left\{ \begin{array}{c}
\omega^{2}\\
1-\omega^{2}\end{array}\right\} \sqrt{\frac{v_{g}}{L|u_{2}|}}(\bar{A}^{+}e^{ix+i\omega t}+\bar{A}^{-}e^{-ix+i\omega t})+\:\mbox{c.c.}\:+\dots,\label{EPA+-}\end{equation}
and the remaining MLE coefficient, $\Delta n$, and the dispersion
coefficient of the NLCMEd, $\varepsilon$, are given by \begin{equation}
\Delta n=\frac{v_{g}}{L|w|}\kappa\quad\textrm{and}\quad\varepsilon=-\frac{d}{Lv_{g}},\label{Dndisp}\end{equation}
which can be computed after making use of (\ref{eigv+omega}), (\ref{velgrupo}),
(\ref{dispersion}), (\ref{grating}) and (\ref{nonlinear}). 

We consider only two configurations because the MLE numerical integrations
are rather CPU costly (large system length and very long final integration
time).

\begin{figure*}[!h]
\begin{centering}\includegraphics[clip,scale=0.75]{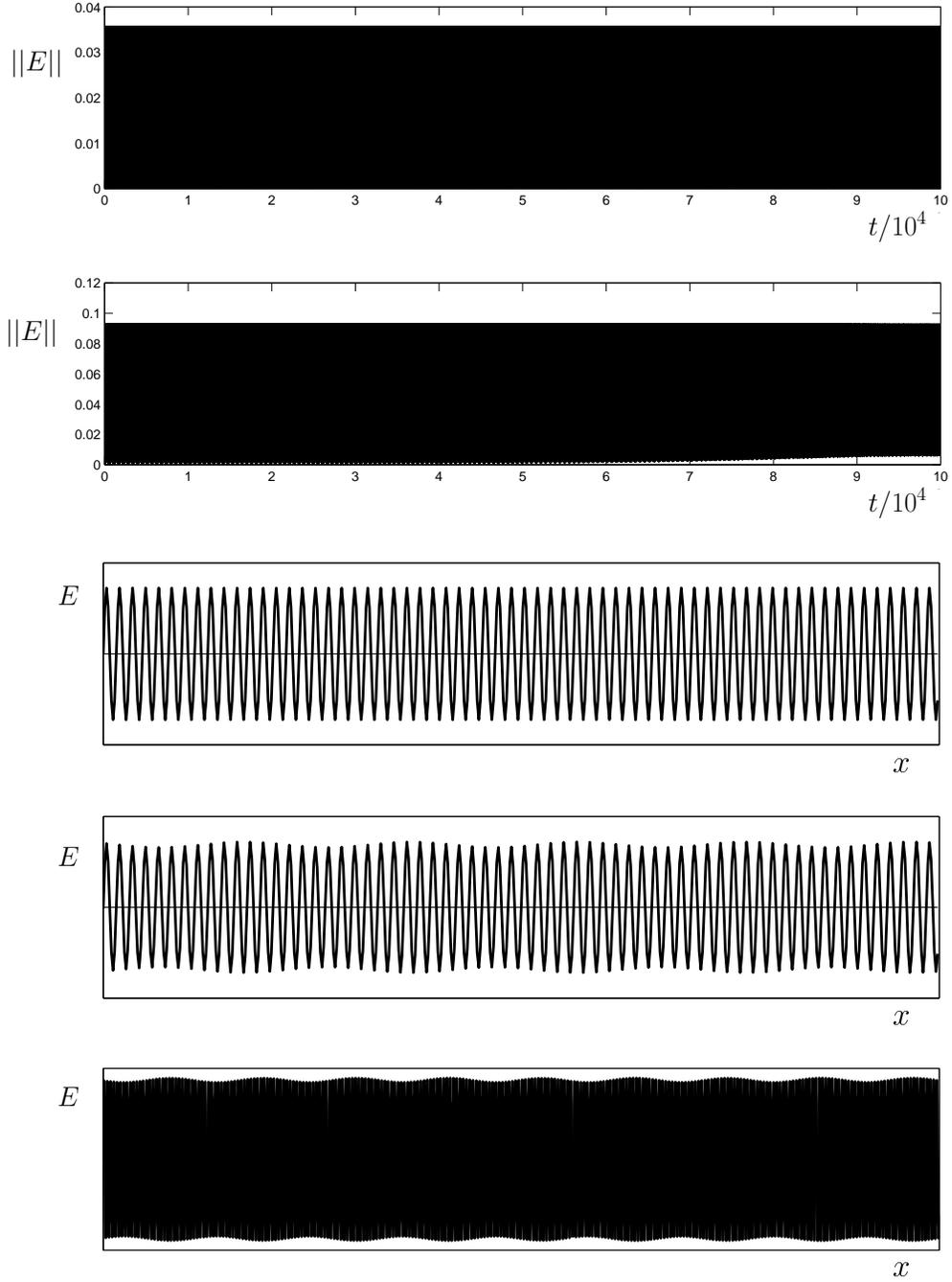}\vspace{-5mm}\par\end{centering}

\caption{\label{fig:CASE1}MLE simulation results starting from a CW ($\kappa=1$,
$\theta=-\frac{\pi}{4}$ and $\rho^{2}=1$) with a $10^{-4}$ perturbation.
From top to bottom: time evolution of the spatial norm of $E$ for
$\omega^{-}$ and $\omega^{+}$, spatial profiles of $E$ at $t=75000$
for $\omega^{-}$ and $\omega^{+}$, and spatial profile of $E$ at
$t=160000$ for $\omega^{+}$ and $L=512\pi$.}
\end{figure*}

\textbf{CASE 1} The initial CW parameters are $\kappa=1$, $\theta=-\frac{\pi}{4}$
and $\rho^{2}=1$. The NLCMEd results presented in ref.~\cite{Martel05}
indicate that this CW is stable for negative dispersion and dispersively
unstable for positive dispersion, while, according to the NLCME, this
CW is always stable. The numerical integrations of the MLE presented
in Fig.~\ref{fig:CASE1} correspond to the parameters $\omega_{p}^{2}=1$,
$n_{0}^{2}=2$, $L=128\pi$ (i.e., there are 128 grating oscillations
inside the fiber ring). The first and second plot correspond, respectively,
to $\omega=\omega^{-}$ and $\omega=\omega^{+}$ (that is, to negative
and positive $\varepsilon$ in the NLCMEd (see eq (\ref{Dndisp})
and Fig..~\ref{fig:disp-rel-detail})) with the MLE grating strength,
$\Delta n$, that results from eq. (\ref{Dndisp}). They show the
time evolution of the spatial norm of the electric field,\[
\| E\|=\sqrt{\frac{1}{L}\int_{0}^{L}|E|^{2}\, dx},\]
and look like a solid black patch due to the fact that $||E||$ oscillates
very fast in time. In agreement with the NLCMEd predictions, the CW
is stable for negative dispersion (first plot in Fig.~\ref{fig:CASE1})
and unstable for positive dispersion (the instability growth can be
appreciated from $t=60000$ on in the second plot of Fig.~\ref{fig:CASE1}).
The corresponding spatial profiles of $E$ at $t=75000$ are given
in the third and fourth plots of Fig.~\ref{fig:CASE1}; for negative
dispersion (third plot) a perfectly uniform amplitude oscillatory
pattern is obtained (the CW pattern) but, for positive dispersion,
a modulation is clearly present (fourth plot). In order to be sure
that this is a dispersive instability we have repeated the unstable
MLE simulation in a four times longer domain ($L=256(2\pi)$). The
resulting spatial profile of $E$ at $t=160000$ is shown in the last
plot of Fig.~\ref{fig:CASE1}. Notice how the number of basic wavelengths
is now four times higher but the number of wavelengths of the modulation
only approximately doubles (increases from 5 to 9), confirming the
dispersive character of the instability whose characteristic size
scale as $\sqrt{L}$ (see ref.~\cite{Martel05}). 

\begin{figure*}[!h]
\begin{centering}\includegraphics[clip,scale=0.75]{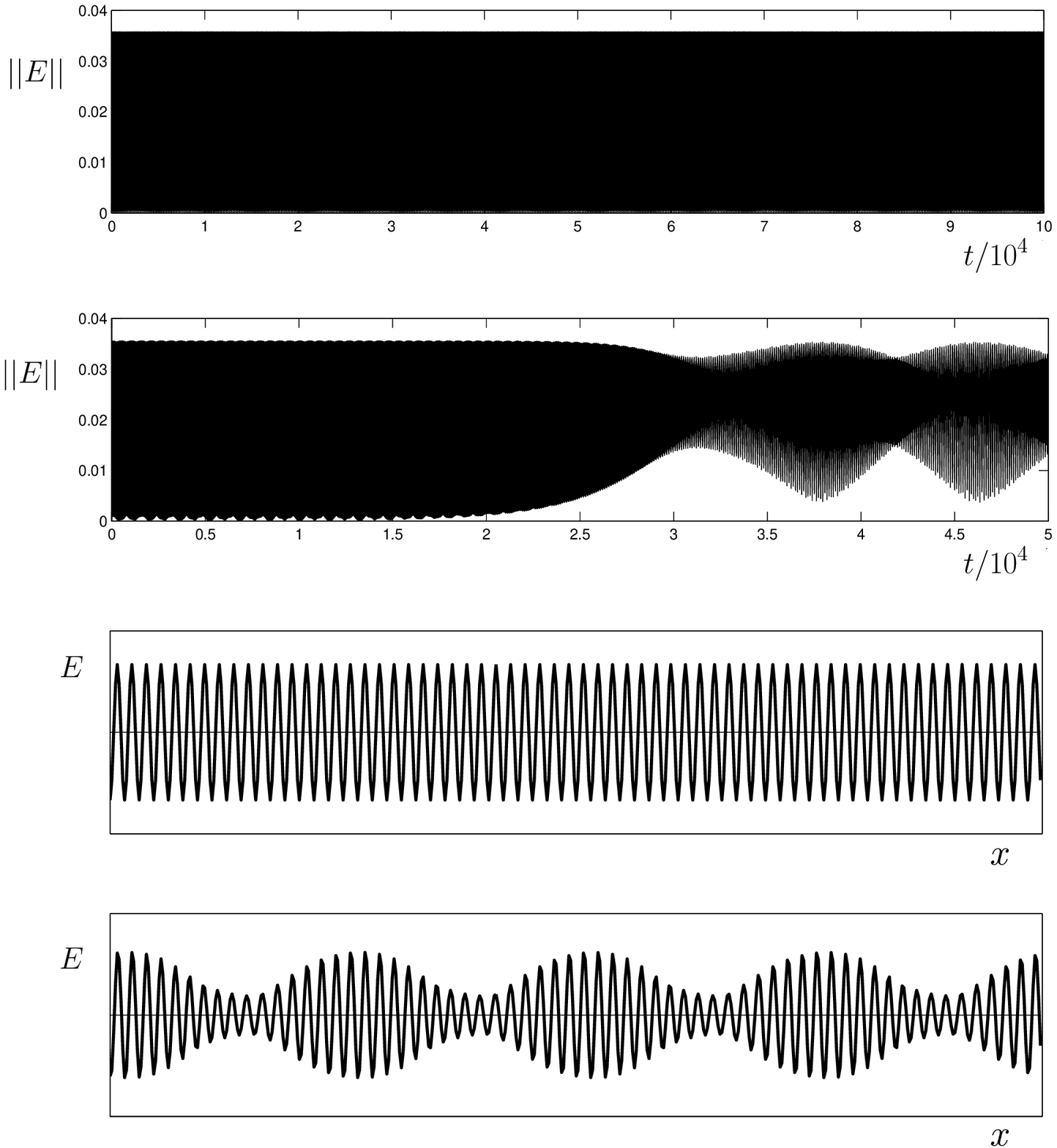}\par\end{centering}

\caption{\label{fig:CASE2} MLE simulation results starting from a CW ($\kappa=1$,
$\theta=\frac{\pi}{4}$ and $\rho^{2}=1$) with a $10^{-4}$ perturbation.
From top to bottom: time evolution of the spatial norm of $E$ for
$\omega^{+}$ and $\omega^{-}$, spatial profile of $E$ at $t=75000$
and $\omega^{+}$, and spatial profile of $E$ at $t=30000$ for $\omega^{-}$.}
\end{figure*}

\textbf{CASE 2} The CW parameters are now $\kappa=1$, $\theta=\frac{\pi}{4}$
and $\rho^{2}=1$, and the MLE parameters are the same as in the above
case: $\omega_{p}^{2}=1$, $n_{0}^{2}=2$ and $L=128\pi$. The first
and third plot of Fig.~\ref{fig:CASE2} correspond to positive dispersion
and indicate that the CW is now stable. The dispersion is negative
in the second and fourth plot where the destabilization of the CW
can be clearly seen both in the time evolution of $||E||$ and in
the dispersive modulations that the spatial profile of $E$ displays.
This is again in perfect agreement with the linear stability results
obtained from the NLCMEd \cite{Martel05} (the NLCME again wrongly
labeled this CW as always stable).

In conclusion, the numerical simulations of the MLE indicate that
the NLCME fail to describe the system evolution if dispersive instabilities
(that cannot be detected using the NLCME formulation) come into play.
In this case the higher order dispersion effects must be taken into
account, and the amplitude equations that do correctly predict the
weakly nonlinear dynamics of light propagation in FBG are the asymptotically
nonuniform NLCMEd (\ref{disp+})-(\ref{disp-}).

\begin{acknowledgments}
This work has been supported by the European Office of Aerospace Research
and Development (FA8655-02-M4087), by the Spanish Dirección General
de Investigación (MTM2004-03808) and by the Universidad Politécnica
de Madrid (R05/11071).\bibliographystyle{apsrev}
\bibliography{CHAOS06}

\end{acknowledgments}

\end{document}